\documentclass{ws-p9-75x6-50}

\begin{document}

\title{Generalized true- and eccentric-anomaly parametrizations for 
 the perturbed Kepler motion}
\author{L\'aszl\'o \'A. Gergely$^{1,2}$, Zolt\'an I. Perj\'es$^1$
and M\'aty\'as Vas\'uth$^1$}
\address{$^1$\ KFKI Research Institute for Particle and Nuclear\\
Physics, Budapest 114, P.O.Box 49, H-1525 Hungary \\
   $^2$\ Astronomical Observatory and Department of Experimental Physics,\\ 
University of Szeged, Szeged, D\'om t\'er 9, H-6720 Hungary  }

\maketitle

\abstracts{The true- and eccentric-anomaly parametrizations of the Kepler
motion are generalized to quasiperiodic orbits by considering perturbations
of the radial part of kinetic energy as a series in the negative powers
of the orbital radius. A toolbox of methods for averaging observables
in terms of the energy $E$ and angular momentum $L$ is developed.
A broad range of systems governed by the generic Brumberg force, as well
as recent applications of the theory of gravitational radiation involve
integrals over a period of motion. These integrals are evaluated by using
the residue theorem. It is shown that the pole of the integrand is
located in the origin and that under certain circumstances an
additional pole emerges. }

In our previous works \cite{GPV3,Laci2} we have analyzed the motion
of a compact binary system with rotating components in order to 
provide realistic signal templates for the interferometric 
gravitational-wave observatories. We have calculated the cumulative
radiation reaction effects on the motion of the spinning binary. 
In obtaining secular changes one has to evaluate integrals of type
\begin{equation} \label{integ}
\int_{0}^{T}\frac{\omega }{r^{2+n}}dt \ ,
\end{equation}
where T is the radial period, n an integer and $\omega$ a function
with no manifest radial dependence. For example $T$ is determined by $n=-2$, while
averaging instantaneous radiative losses of quantities characterizing the 
orbit of a binary system one calculates this type of integrals with
$n=1,\dots,7$. During the computations it is convenient to use the 
generalized true- and eccentric-anomaly parameters defined as follows.

The radial equation of a perturbed Kepler problem is considered as
\begin{equation}
\dot{r}^{2}=\frac{2E}{\mu }+\frac{2m}{r}-\frac{L^{2}}{\mu ^{2}r^{2}}
+\sum_{i=0}^{p}\frac{\varphi _{i}}{\mu ^{2}r^{i}} \ , \label{rdot}
\end{equation}
where the small coefficients $\varphi _{i}$ couple the perturbations
to the system. The conservation of energy $E$ and magnitude of the
orbital angular momentum $L$ are assumed to hold from symmetry 
considerations. A wide class of perturbed Kepler motions 
accomplish these conditions, for instance systems described by
the Brumberg\cite{brum} force and the dynamics of spinning binaries.

The generalized true- and eccentric-anomaly parametrizations are 
introduced by the relations
\begin{equation}
\frac{dr}{d(\cos \chi )}=-\gamma r^{2}\ , \quad
\frac{dr}{d(\cos \xi  )}=-\kappa   \ ,     
\end{equation}
where $\gamma$ and $\kappa$ are positive constants.
The scale of the parameters is adjusted to the turning points
$r(0)=r_{\min}$, $r(\pi )=r_{\max}$, which are solutions 
of the radial equation $\dot{r}=0$. 
To first order one has $p-2$ unphysical roots (even if real, 
they are much smaller than the radius $r$) and $r_{\min}$ and $r_{\max}$,
which differ from the Keplerian turning points in linear terms in
$\varphi_i$. With the above properties the familiar form of the
parametrizations is obtained
:
\begin{equation}                                                                                                    
\frac{2}{r}=\frac{1+\cos \chi }{r_{\min}}+\frac{1-\cos \chi }{r_{\max}}         
\ , \quad                                                                 
2r=\frac{1+\cos \xi}{r_{\min}^{-1}}+\frac{1-\cos \xi}{r_{\max}^{-1}} \ .           
\end{equation}     

The most remarkable properties of these parametrizations are summarized 
in the following theorems:

{\bf Theorem 1}: {\sl The integral }$\int r^{-(2+n)}dt${\sl \ , }$
n\geq 0${\sl \ is given by the residue located in the origin of the
plane of the complex
 true-anomaly parameter} $z=e^{i\chi }${\sl . }

{\bf Theorem 2}: {\sl The integral }$\int r^{-(2+n)}dt$ with $n<0$
{\sl \ is given by the sum of the residues in the origin of the complex
parameter plane $w=e^{i\xi }$ and [for perturbative terms with 
powers $i\ge 2-n$ of $r$ in the radial equation (\ref{rdot})] at}
\begin{eqnarray}
w_{1}=\left( \frac{m\mu ^{2}-\sqrt{-2\mu EL^{2}}}{m\mu ^{2}+\sqrt{-2\mu
EL^{2}}}\right) ^{1/2}\ . \nonumber
\end{eqnarray}

The above statements are proved by passing to the true- and 
eccentric-anomaly parameters in the integral (\ref{integ})
and examining this expression in linear order of the perturbations
$\varphi_i$. In doing so one can specify the possible $\omega$ 
functions which will not destroy the advantageous property of the 
integrand. It turns out, that $\omega$ can be any polynomial in
sine and cosine of the parameters. In particular, it can have the
form 
\begin{equation}
 \label{omeg}
\omega =\omega _{0}+\varphi (\psi ,\dot{r}) \ ,
\end{equation}
where $\omega_0$ is constant and the perturbing term
$\varphi (\psi ,\dot{r})$ is a polynomial in $\sin\psi$, $\cos\psi$
and $\dot{r}$. To first order the expressions of the azimuthal
angle $\psi$ and $\dot{r}$ in terms of the true- and 
eccentric-anomaly parameters are needed to Keplerian order.

The two types of generalized Keplerian parametrizations studied 
here are capable of describing a wide variety of systems (\ref{rdot}).
They do not give a complete parametrization of the motion, but are suitable
for handling radial effects. When using them and the residue theorem
the evaluation of various integrals (\ref{integ}) turns out to
be a simple task.
 They have been tested in computing the secular
 radiative
changes of a compact binary system\cite{GPV3,Laci2}. In our view the solution 
of the radial motion presented here is at least a viable candidate
amongst other approaches \cite{RS,GI}.


\begin{thebibliography}{99}


\bibitem{GPV3}  L.Gergely, Z.Perj\'{e}s and M.Vas\'{u}th,
                  \Journal{\PRD}{58}{124001}{1998}.


\bibitem{Laci2}  L.\'{A}.Gergely, \Journal{\PRD}{61}{024035}{2000}.

\bibitem{brum}  Brumberg,V.A.: Essential Relativistic Celestial Mechanics,
Bristol: Adam Hilger, 1991.

\bibitem{RS}  R. Rieth and G. Sch\"afer, {\em Class. Quantum Grav.}
     \ {\bf 14}, 2357 (1997).

\bibitem{GI}  Gopakumar,A. and Iyer,B.R., \Journal{\PRD}{56}{7708}{1997}.

\end{thebibliography}
\end{document}